\documentstyle[aps,prb,multicol,eqsecnum]{revtex}

\begin{document}

\title{Perturbation study on the spin and charge
  susceptibilities \\ of the two-dimensional Hubbard model}

\author{Takashi Hotta}
\address{Institute for Solid State Physics, University of Tokyo,
  7-22-1 Roppongi, Minato-ku, Tokyo 106, Japan}

\author{Satoshi Fujimoto}
\address{Department of Physics, Kyoto University, Kyoto 606, Japan}

\date{\today}

\maketitle

\begin{abstract}

We investigate the spin and charge susceptibilities of the 
two-dimensional Hubbard model based upon the perturbative
calculation in the strength of correlation $U$. 
For $U$ comparable to a bare bandwidth, the charge susceptibility 
decreases near the half-filling as hole-doping approaches zero. 
This behavior suggesting the precursor of the Mott-Hubbard gap 
formation cannot be obtained without the vertex corrections beyond the 
random phase approximation. 
In the low-temperature region, the spin susceptibility deviates 
from the Curie-Weiss-like law and finally turns to decrease with the
decrease of temperature. 
This spin-gap-like behavior is originating from the 
van Hove singularity in the density of states. 

\end{abstract}


\begin{multicols}{2}

\section{Introduction}

The Hubbard model \cite{hubbard} has been studied extensively as the 
fundamental model for strongly correlated electron systems which
exhibit the Mott-Hubbard metal-insulator transition (MIT) \cite{mott}.
Especially, the two-dimensional Hubbard model (2DHM) has been 
attracted much attention in relation to high-temperature 
superconductors (HTSC) of which normal metallic phase is quite
``anomalous'' in the sense that its property seriously deviates from 
that predicted by the conventional Fermi-liquid theory. 
Thus in this paper, we investigate 2DHM putting stress on the
correlation effects on the charge and spin susceptibilities and 
grasp the signal of the ``anomalous'' behavior in the charge and 
spin responses near MIT from the weak-coupling perturbation approach. 

As for the charge degree of freedom, we are interested in the 
dependence on hole-doping $\delta$ of the charge susceptibility. 
Since the system becomes incompressible near MIT, it is generally 
expected that the charge susceptibility decreases and eventually
vanishes when we approach MIT by putting $\delta$ as zero. 
In fact, such a behavior has been obtained in some numerical studies
on the basis of the quantum Monte Carlo (QMC) method. \cite{moreo,oht} 
A quite different result, however, has been reported in QMC
simulations \cite{imada}: The charge susceptibility diverges like 
$\sim 1/\delta$ as $\delta$ approaches zero. 
The difference between two results is not brought about by the scatter 
of the QMC data among the groups, but is originating from the physical 
interpretation of data.  
Thus it is desirable to investigate the doping dependence of the
charge susceptibility with use of other methods. 
In this paper, by using the second-order perturbation theory (SOPT)
with respect to the on-site Coulomb interaction $U$, we investigate
the tendency of the change of the charge susceptibility from that in
the non-interacting case due to correlation effects. 

On the other hand, as for an anomalous behavior of the spin response, 
we would like to note the spin-gap-like behavior observed in the
nuclear spin relaxation rate $T_1^{-1}$ of several HTSC materials. 
\cite{kitaoka1,kitaoka2,yasuoka1,warren,takigawa,zimm,yasuoka2,zheng,wals} 
In the higher-temperature region than the superconducting transition 
temperature $T_c$, $(T_1T)^{-1}$ increases in proportional to
$1/(T+\theta)$ with the decrease of $T$, indicating that $(T_1T)^{-1}$ 
obeys the Curie-Weiss law. Here $\theta$ is the Weiss temperature. 
As $T$ is decreased, in such materials as 
La$_{1-x}$Sr$_x$CuO$_4$ \cite{kitaoka1} and 
YBa$_2$Cu$_3$O$_{7-\delta}$ with $T_c \sim 90$K,
\cite{kitaoka2,yasuoka1} 
$(T_1T)^{-1}$ continues to increase untill there occurs the transition
to superconducting state. 
However, in such materials as 
YBa$_2$Cu$_3$O$_{6.6}$ with $T_c \sim 60$K,
\cite{yasuoka1,warren,takigawa}
YBa$_2$Cu$_4$O$_8$, \cite{zimm,yasuoka2,zheng} 
and Bi$_2$Sr$_2$CaCuO$_8$, \cite{wals} 
$(T_1T)^{-1}$ begins to deviate from the Curie-Weiss law even for 
$T > T_c$, and turns to decrease with the decrease of $T$. 
Since this decrease of $(T_1T)^{-1}$ seems to be associated with 
the formation of the spin-singlet state, 
such a behavior has been frequently called ``the spin gap''. 
Some authors have argued the origin of the spin-gap formation such as 
the spinon pairing \cite{fuku} or bi-layer coupling, \cite{millis} 
but it should be clarified whether it is possible or not to reproduce 
the spin-gap behavior in the framework of the Fermi-liquid theory
without bi-layer coupling. From the analysis of the normal state 
of HTSC on the basis of the 
Fermi-liquid theory in due consideration of anti-ferromagnetic (AF)
spin fluctuations, \cite{kohno} the following relation holds in HTSC: 
\begin{equation}
  (T_1T)^{-1} \propto \chi_s({\bf Q},0), 
\end{equation}
where $\chi_s({\bf Q},0)$ is the staggered spin susceptibility in the 
static limit with ${\bf Q}=(\pi,\pi)$. 
Thus in this paper, by calculating $\chi_s({\bf Q},0)$ in SOPT, 
we make a qualitative investigation on the temperature dependence of
$(T_1T)^{-1}$ of 2DHM. 

By using the perturbative calculation, several authors have studied
2DHM so far, \cite{sch,hot1,zlatic} but the spin and charge
susceptibilities which reflect the correlation 
effects on the spin and charge excitation have not been studied well 
even from the view point of SOPT. 
Although the perturbation method is not appropriate for the
description of phase transition such as MIT, it is useful to study how 
correlation effects modify the behavior of these quantities in the 
Fermi-liquid phase and cause ``anomalous'' behaviors. 
We can point out such a merit of the perturbative approach that it is
free from size effects compared with numerical simulations and 
suitable for the description of the low-energy properties. 
On the other hand, there exists such a demerit that the perturbative 
calculation will fail down for enough large value of $U$. 
However, as is seen in the perturbation study of the single impurity
Anderson model, the expansion in terms of $U$ is expected to converge
asymptotically unless the spontaneous symmetry breaking occurs,
leading to that the truncation of the expansion in second order may
give a qualitatively correct description of some physical properties. 
Moreover, SOPT is helpful to grasp some features in the limit of
large value of $U$. 
For example, it properly predicts the location of the Mott-Hubbard
band at half-filling. 
In this sense, we may expect that our results on the spin and charge
susceptibilities are qualitatively valid even for large value of $U$
near half-filling. 
Thus it is meaningful to study the strong correlation effects within 
SOPT.   

In order to evaluate the spin and charge susceptibilities properly in
the vicinity of the half-filling, we take into account the vertex
corrections which are not included in the random phase approximation
(RPA). 
These corrections are considered to be important in two-dimensional 
systems, because the charge and spin fluctuations which destroy 
long-range order are large and a simple mean-field description such as 
RPA will not be valid. 
Actually, the vertex corrections beyond RPA are important for the 
description of the charge excitations near the half-filling as we will
see later. 
The calculation is performed in the framework of the well-known
Green's function formalism. 
In order to carry out the complicated numerical computation, 
we exploit the Fast-Fourier-Transformation (FFT) algorithm. 

The organization of this paper is as follows. 
In Sec.~II, we develop the formulation to perform our calculation 
on the basis of the microscopic Fermi-liquid theory. 
In Sec.~III, the numerical results for the charge susceptibility are 
given. 
We investigate the usual Hubbard model only with nearest-neighbor 
hopping terms as well as the model with next nearest-neighbor hopping 
ones which deforms the shape of the Fermi surface and changes the 
location of the van Hove singularity (VHS). 
For $U$ comparable to the nearest-neighbor hopping energy, the charge 
susceptibility decreases as $\delta$ approaches zero 
due to the suppression effect of the vertex corrections. 
In Sec.~IV, we show our numerical results for the dependence on 
$\delta$ as well as $T$ of the spin susceptibility. 
In the low-temperature region, the spin susceptibility shows
spin-gap-like behavior due to VHS in the density of states (DOS). 
Finally in Sec.~V, we summarize this paper and briefly comment on the
relation between our results and NMR experimental results in HTSC. 
We will employ units in which both $\hbar$ and $k_{\rm B}$ are taken 
to unity.

\section{Formulation}

\subsection{Model Hamiltonian}

Let us consider tight-binding electrons on the two-dimensional square 
lattice. 
The non-interacting part of the Hamiltonian consists of the hopping 
term between nearest- and next nearest-neighbor sites. 
We take into account the on-site Coulomb repulsion $U$. 
Thus, we obtain the following model Hamiltonian: 
\begin{eqnarray}
  \label{hamiltonian}
  H &=& \sum_{{\bf k}\sigma} (E_{\bf k}-\mu) 
  c^{\dagger}_{{\bf k}\sigma}c_{{\bf k}\sigma} \nonumber \\ 
  &+& (U/N) \sum_{{\bf k,k'}}\sum_{{\bf q}(\ne 0)}
  c^{\dagger}_{{\bf k-q}\uparrow}c^{\dagger}_{{\bf k'+q}\downarrow}
  c_{{\bf k'}\downarrow} c_{{\bf k}\uparrow}, 
\end{eqnarray}
where $c_{{\bf k}\sigma}$ is the annihilation operator of an electron 
with momentum ${\bf k}$ and spin $\sigma$, $\mu$ the chemical
potential, $U$ the on-site Coulomb interaction and $N$ the number of 
sites. 
The electronic dispersion relation is given by 
\begin{equation}
  \label{dispersion}
  E_{\bf k}=-t(\cos k_x + \cos k_y)-t' \cos k_x \cos k_y, 
\end{equation}
where $t$ and $t'$ are, respectively, proportional to nearest and
next-nearest neighbor hopping integrals. 

\subsection{Self-energy in SOPT}

In general, the one-electron Green's function $G(k)$ is expressed as 
\begin{equation}
  \label{gfn}
  G(k)=\frac{1}{i\epsilon_n+\mu-E_{\bf k}-\Sigma(k)}, 
\end{equation}
where $\Sigma(k)$ is the electronic self-energy, 
$k$ denotes $({\bf k},i\epsilon_n)$, 
$\epsilon_n \equiv \pi T (2n+1)$ with an integer $n$ is the fermion 
Matsubara frequency at a temperature $T$, and $\mu$ is determined
through the relation 
\begin{equation}
  \label{number}
  n=2\sum_k G(k) e^{i\epsilon_n \eta}. 
\end{equation}
Here $n$ is the number density of electron, $\eta$ is the positive
infinitesimal, and we use such a shorthand notation as  
\begin{equation}
  \sum_k \equiv T \sum_n \sum_{\bf k}. 
\end{equation}
With use of $n$, the hole-doping $\delta$ is defined as 
\begin{equation}
  \delta \equiv 1-n. 
\end{equation}

The Hartree term of the self-energy is expressed as 
\begin{equation}
  \label{hartree}
  \Sigma_{\rm H}=U \sum_k G(k) e^{i\epsilon_n \eta}, 
\end{equation}
which is equal to $Un/2$, leading to a constant energy shift. 
Thus, we define the Green's function in the Hartree approximation, 
$G_0(k)$, as 
\begin{equation}
  \label{gfn0}
  G_0(k)=\frac{1}{i\epsilon_n+\mu_0-E_{\bf k}}, 
\end{equation}
where $\mu_0$ is determined through the relation Eq.~(\ref{number}) 
in which $G$ is replaced with $G_0$. 
In this number-conserving approximation, 
the Hartree term is automatically taken into
account in $G_0$. 
Therefore in the following, we do not include the Hartree term 
explicitly in each order term. 

Since the Hartree term is absorbed into $\mu_0$, we need to calculate
only the second-order term of the self-energy, given by 
\begin{equation}
  \label{self}
  \Sigma(k)=U^2 \sum _{k'} \chi_0(k-k') G_0(k'), 
\end{equation}
where $\chi_0(q)$ is defined by
\begin{equation}
  \label{chi0}
  \chi_0(q)=-\sum_k G_0(k) G_0(k+q). 
\end{equation}
Here we use another shorthand notation as 
$q = ({\bf q},i\omega_{\nu})$, where 
$\omega_{\nu} \equiv 2\pi T \nu$ with an integer $\nu$ is 
the boson Matsubara frequency. 

Let us evaluate $\mu$ within SOPT, expressed as 
\begin{equation}
  \label{chemi}
  \mu=\mu_0+\delta \mu, 
\end{equation}
where $\delta \mu$ is the chemical potential due to the second-order 
term. From Eqs.~(\ref{gfn}), (\ref{gfn0}), (\ref{self}), and
(\ref{chemi}), up to second order in $U$, $n$ in Eq.~(\ref{number}) 
is expressed as 
\begin{equation}
  n= 2\sum_k G_0(k) e^{i\epsilon_n \eta} + 2\sum_k G_0^2(k) 
  \bigl[ \Sigma(k) -\delta \mu \bigr]. 
\end{equation}
Since the electron number is conserved, $\delta \mu$ is given by 
\begin{equation}
  \delta \mu = -\frac{1}{\chi_0(0)} \sum_k G^2_0(k) \Sigma(k). 
\end{equation}
In the following, we rewrite $\Sigma(k)-\delta \mu$ as $\Sigma(k)$. 

\subsection{Charge and spin susceptibilities in SOPT}

We define $\chi_s(q)$ and $\chi_c(q)$ as 
\begin{equation}
  \label{def:chic}
  \chi_c(q) = 
  \int_0^{1/T} d\tau e^{i\omega_{\nu} \tau} 
  \bigl\langle T_{\tau} [~\Delta n_{\bf q}(\tau) \Delta n_{\bf -q}~] 
  \bigr\rangle,
\end{equation}
and 
\begin{equation}
  \label{def:chis}
  \chi_s(q) = (g\mu_B)^2
  \int_0^{1/T} d\tau e^{i\omega_{\nu} \tau}
  \bigl\langle T_{\tau} [~S_{\bf q}^z (\tau)
  S_{-{\bf q}}^z~] \bigr\rangle, 
\end{equation}
where $S_{\bf q}^z$ and $\Delta n_{\bf q}$ are, respectively, given by 
\begin{equation}
  S_{\bf q}^z = (1/2) \sum_{\bf k}
  [c^{\dagger}_{{\bf k+q}\uparrow}c_{{\bf k}\uparrow}
  -c^{\dagger}_{{\bf k+q}\downarrow}c_{{\bf k}\downarrow}], 
\end{equation}
and $\Delta n_{\bf q}=n_{\bf q}-\langle n_{\bf q} \rangle$ with 
\begin{equation}
  n_{\bf q}=\sum_{{\bf k}\sigma}
  c^{\dagger}_{\bf{k+q}\sigma}c_{{\bf k}\sigma}. 
\end{equation}
Here $g$ is the electron $g$-factor and $\mu_{B}$ is the Bohr
magneton. The magnetic unit is fixed as $g\mu_B/2=1$. 

Equations (\ref{def:chic}) and (\ref{def:chis}) can be recast into 
\begin{equation}
  \label{eq:chic}
  \chi_c(q) = 2 [\chi^{\uparrow \uparrow}(q) 
  + \chi^{\uparrow \downarrow}(q)],
\end{equation}
and
\begin{equation}
  \label{eq:chis}
  \chi_s(q) = 2 [\chi^{\uparrow \uparrow}(q) 
  - \chi^{\uparrow \downarrow}(q)],
\end{equation}
respectively, where $\chi^{\sigma \sigma'}(q)$ is defined as 
\begin{eqnarray}
  \label{def:chi}
  \chi^{\sigma \sigma'}(q) &=&
  \int_0^{1/T} d\tau e^{i\omega_{\nu} \tau} 
  \bigl\langle T_{\tau} \sum_{\bf k}
  c^{\dagger}_{\bf{k+q}\sigma}(\tau) 
  c_{{\bf k}\sigma}(\tau) \nonumber \\ 
  &\times& \sum_{\bf p} c^{\dagger}_{\bf{p+q}\sigma}
  c_{{\bf p}\sigma} \bigr\rangle_{\rm conn}, 
\end{eqnarray}
where the subscript ``conn'' denotes the operation to take into
account only connected diagrams. 
Expanding the right-hand side of Eq.~(\ref{def:chi}) with respect to
$U$ up to second order in the framework of the conventional
perturbation theory, we obtain $\chi^{\sigma \sigma'}(q)$
diagrammatically shown in Fig.~\ref{fig:diagrams}. 
Combining Eqs.~(\ref{eq:chic}) and (\ref{eq:chis}) with the explicit 
expressions for $\chi^{\sigma \sigma'}(q)$, 
we obtain $\chi_s(q)$ and $\chi_s(q)$ as 
\begin{eqnarray}
  \chi_c(q) &=& 2 \bigl[ \chi_0(q)-U\chi_0^2(q)+U^2\chi_0^3(q) 
  \nonumber \\ 
  &-& \delta \chi(q)-U^2 \{ 2 \chi_{\rm corr1}(q)
  +\chi_{\rm corr2}(q) \} \bigr], 
\end{eqnarray}
and 
\begin{eqnarray}
  \chi_s(q) &=& 2 \bigl[ \chi_0(q)+U\chi_0^2(q)+U^2\chi_0^3(q) 
  \nonumber \\ 
  &-& \delta\chi(q)+U^2 \chi_{\rm corr2}(q) \bigr], 
\end{eqnarray}
where both $\chi_{\rm corr1}(q)$ and $\chi_{\rm corr2}(q)$ indicate 
the vertex corrections which are not included in RPA and
$\delta\chi(q)$ is the correction term due to the insertion of the 
self-energy. 
The expressions for $\chi_{\rm corr1}(q)$, $\chi_{\rm corr2}(q)$, 
and $\delta\chi(q)$ are given in the following: 
\begin{eqnarray}
  \label{corr1}
  \chi_{\rm corr1}(q)&=&\sum_{k_1,k_2} 
  G_0(k_1) G_0(k_1+q) \chi_0(k_1-k_2) \nonumber \\
  &\times& G_0(k_2) G_0(k_2+q), 
\end{eqnarray}
\begin{eqnarray}
  \label{corr2}
  \chi_{\rm corr2}(q) &=& \sum_{k_1,k_2}
  G_0(k_1) G_0(k_1+q) \phi_0(k_1+k_2) \nonumber \\ 
  &\times& G_0(k_2) G_0(k_2-q),
\end{eqnarray}
with $\phi_0(q)$ defined by 
\begin{equation}
  \phi_0 (q) \equiv -\sum_k G_0(k) G_0(q-k), 
\end{equation}
and 
\begin{eqnarray}
  \delta \chi(q)=\sum_{k} G_0^2(k) \Sigma(k) 
  \bigl[ G_0(k+q)+G_0(k-q) \bigr]. 
\end{eqnarray}
For the comparison, we also consider the susceptibilities calculated
in RPA-like approximation up to second order in $U$, 
which are given by
\begin{equation}
  \chi_c^{\rm RPA}(q)=2[\chi_0(q)-U\chi_0^2(q)+U^2\chi_0^3(q)], 
\end{equation}
and
\begin{equation}
  \chi_s^{\rm RPA}(q)=2[\chi_0(q)+U\chi_0^2(q)+U^2\chi_0^3(q)]. 
\end{equation}
In the following, we call this approximation 
``second-order RPA (SO-RPA)''. 

\subsection{Method for numerical calculation}

When we carry out sum over the momentum as well as the Matsubara 
frequency in the numerical calculation, the first Brillouin zone (FBZ)
is divided into fine meshes and the frequency sum is terminated at a
cut-off energy $\omega_c$. 
In this paper, we perform the calculation on a $64 \times 64$ lattice 
from the limitation of computer memory. 
We should note an existence of a lower limit of temperature in which
reliable results are obtained: 
As has been pointed out by Serene and Hess \cite{fft2}, with a 
$64 \times 64$ lattice, the deviation from the Fermi-liquid behavior
becomes significant at $T/t=0.0075$ for our definition of $t$. 
Thus, our calculation will be carried out only for $T$ larger than 
$0.01t$. 

As for the frequency cut-off, we always choose $\omega_c$ as $64t$, 
irrespective of $T$. 
Note that $\omega_c$ is equal to $16W$, where $W$ is a bare bandwidth
given by $4t$. 
Such a large value for the cut-off energy is considered to be
sufficient to converge the frequency sum. \cite{fft2}

We briefly explain a means to perform the complicated sum 
over $k$ efficiently. 
The point is to arrange the sum into the convolution form convenient 
for the FFT algorithm. \cite{fft1,fft2} 
For example, $\chi_{\rm corr1}$ in Eq.~(\ref{corr1}) is rewritten as 
\begin{equation}
  \chi_{\rm corr1}(q)=\sum_{k_1,k_2}
  C_1(k_1;q) \chi_0(k_1-k_2) C_1(k_2;q),
\end{equation}
where $C_1(k;q)=G_0(k) G_0(k+q)$. 
When we symbolically represent the Fourier transform of 
$\chi_0(q)$ and $C_1(k;q)$ as 
\begin{equation}
  \label{eq:fft1}
  \chi_0(q)=\sum_x e^{iqx}\chi_0(x),
\end{equation}
and 
\begin{equation}
  \label{eq:fft2}
  C_1(k;q)=\sum_y e^{iky}C_1(y;q),
\end{equation}
respectively, $\chi_{\rm corr1}$ is expressed as 
\begin{equation}
  \label{corr1fft}
  \chi_{\rm corr1}(q) \propto \sum_x C_1(x;q) \chi_0(x) C_1(-x;q), 
\end{equation}
except for the normalization. 
Let us estimate the numerical steps necessary for the calculation with 
and without FFT. 
Assume that we need $M$ steps to carry out sum over $k$. 
When we calculate $\chi_{\rm corr1}$ in Eq.~(\ref{corr1}) without 
FFT for a fixed $q$, the steps of the order of $M^2$ are necessary. 
On the other hand, if we exploit the FFT algorithm for the calculation 
of $\chi_{\rm corr1}$ in Eq.~(\ref{corr1fft}), we need only the steps
of the order of $M{\rm log}_2M$ to finish all the sums. 
This difference leads us a drastic reduction of the task in the
numerical calculation. 

\section{Calculated results for charge susceptibility}

\subsection{Doping dependence}

Now we investigate the dependence on $\delta$ of the uniform charge 
susceptibility in the static limit, $\chi_c$. 
We first consider the case only with nearest-neighbor hopping in the 
dispersion relation given by Eq.~(\ref{dispersion}). 
The results for this case are shown in Fig.~\ref{fig:chicdoping1}. 
For $U=0$, $\chi_c$ increases logarithmically due to VHS, as $\delta$
approaches zero. 
In the interacting case, $\chi_c$ is totally suppressed both in SO-RPA
and SOPT. 
For $U$ comparable to $t$, $\chi_c$ still increases with the decrease
of $\delta$ in SO-RPA, but in contrast, it turns to decrease near the
half-filling in SOPT as $\delta$ approaches zero. 
The large suppression of $\chi_c$ in SOPT near the half-filling is due
to the vertex corrections which are not included in RPA. 

In order to discuss the physical meaning of our calculated results, we 
would like to comment on the relation between the formula for $\chi_c$ 
of the Landau's Fermi-liquid theory and one of the microscopic
Fermi-liquid theory developed by Luttinger. \cite{lutt}
In the Landau's Fermi-liquid theory, $\chi_c$ is phenomenologically
expressed as 
\begin{equation}
  \label{eqn:chila}
  \chi_c=\frac{m^{*}}{m}\frac{\chi_c^0}{1+F_0^s},
\end{equation}
where $m^{*}$ is the renormalized mass, $\chi_c^0$ the bare
susceptibility, and $F^s_0$ the Landau's parameter. 
Based upon the Luttinger's formulation, we obtain another expression
for $\chi_c$ as 
\begin{equation}
  \label{eqn:chicl}
  \chi_c=\sum_{\bf k} \delta (E_{\bf k}^{*}) z_{\bf k}
  \biggl[ 1-\frac{\partial {\rm Re}\Sigma({\bf k},0)}
  {\partial \mu} \biggr], 
\end{equation}
where $E_{\bf k}^{*}$ is the dispersion relation for the
quasi-particle given by 
\begin{equation}
  E_{\bf k}^{*} = z_{\bf k} [E_{\bf k}-\mu+{\rm Re}\Sigma({\bf k},0)], 
\end{equation}
and $z_{\bf k}$ is the renormalization factor determined by 
\begin{equation}
  z_{\bf k}^{-1} = 1- 
  \frac{\partial {\rm Re}\Sigma({\bf k},\omega)}{\partial \omega}
  \biggl\vert_{\omega=0}. 
\end{equation}
Comparing Eq.~(\ref{eqn:chicl}) with Eq.~(\ref{eqn:chila}), we notice
approximately the following correspondence: 
\begin{equation}
  \frac{m^{*}}{m} \chi^0_s \quad \leftrightarrow \quad 
  \sum_{\bf k} \delta (E_{\bf k}^{*}), 
\end{equation}
and
\begin{equation}
  \frac{1}{1+F_0^s} \quad \leftrightarrow \quad 
  z_{\bf k} \biggl[ 1-\frac{\partial {\rm Re}\Sigma({\bf k},0)}
  {\partial\mu} \biggr].
\end{equation}
\noindent From Eq.~(\ref{eqn:chicl}), we see that $\chi_c$ is nothing 
but the average over the Fermi surface of the factor                     
$1-\partial {\rm Re}\Sigma /\partial \mu$. 
It does not depend explicitly on the renormalized mass in contrast to 
Eq.~(\ref{eqn:chila}), but is determined by the bare DOS and the
factor $1-\partial {\rm Re}\Sigma /\partial \mu$ which is incorporated
into our calculation as the vertex corrections. 
Thus the correlation effect on $\chi_c$ is essentially due to the
factor $1-\partial {\rm Re}\Sigma /\partial \mu$. 
Note that $\chi_c$ is not enhanced by the renormalized mass $m^{*}$ 
as naively expected from Eq.~(\ref{eqn:chila}).
This is contrasted with the one-dimensional Luttinger liquid where the 
charge susceptibility is determined solely by the renormalized 
DOS. \cite{lutchi}

Our numerical results imply that 
$1-\partial {\rm Re} \Sigma /\partial \mu$  tends to be zero 
as $\mu$ approaches the half-filling value, 
suggesting that $\chi_c$ vanishes continuously if we can describe 
the MIT transition properly. 
Our results are consistent with that of the QMC simulations by Moreo 
{\it et. al.} \cite{moreo} and Otsuka \cite{oht} but not with the 
assertion by Furukawa and Imada \cite{imada} that the charge 
susceptibility diverges in the vicinity of the half-filling. 
In the framework of the Fermi-liquid theory, the increase of the 
charge susceptibility due to correlation effects does not occur as 
discussed above. 
Thus the conclusion in Ref.~5 implies that some 
non-Fermi-liquid state realizes in the metallic phase near MIT. 
However, the results in the infinite-dimensional Hubbard model
indicate that there is no intermediate exotic state between 
the Fermi-liquid metallic phase and the Mott insulating phase. 
\cite{infinite} 
Since no symmetry breaking occurs at finite temperature in 
two-dimensional systems, we expect that there is no phase transition 
from the Fermi-liquid state to a non-Fermi-liquid one as $n$
approaches unity. 
Thus our results based upon the Fermi-liquid theory plausibly 
may describe the qualitative behavior of the charge 
susceptibility near MIT. 

\subsection{Effects of next-nearest neighbor hopping} 

Since $\chi_c$ depends on the bare DOS, it is sensitively affected by
the band structure. 
In this subsection, we consider the effect of the next-nearest hopping
term which changes the location of VHS and deforms the shape of the
Fermi surface. 
The results for the case of $t'/t=0.2$ is shown in 
Fig.~\ref{fig:chicdoping2}. 
It can be seen that $\chi_c$ increases monotonically as $\delta$
approaches zero, indicating that we cannot obtain the suppression 
effect large enough to overcome the enhancement of the bare DOS
due to VHS. 
In this case, VHS is located in the region of $\omega > 0$, where 
$\omega$ is the energy measured from the Fermi level. 
Thus the bare DOS at the Fermi level is always too small to produce 
strong correlation effects through the vertex corrections 
when we decrease hole-doping by changing the chemical potential. 
The suppression effect due to the vertex corrections is weak 
when the bare DOS at the Fermi level is not large enough. 
This tendency can be confirmed from the results for the case of
$t'/t=-0.2$, which are shown in Fig.~\ref{fig:chicdoping3}. 
In this case, VHS is located in the region of $\omega < 0$. 
When $\delta$ is decreased with the change of $\mu$, the Fermi level 
inevitably crosses the position of VHS at some value of $\mu$. 
At the corresponding hole doping, $\chi_c$ shows a shallow minimum 
for $U/t=1.0$. 
This minimum is due to the maximum of the suppression effect of the 
vertex corrections. 
Thus we conclude that the large bare DOS at the Fermi level results 
in the strong suppression effect of $\chi_c$ through the vertex
corrections which overcomes the enhancement effect of $\chi_c$ 
due to the bare DOS, leading to the total decrease of $\chi_c$ 
in the vicinity of MIT. 

\section{Calculated results for spin susceptibility}

\subsection{Doping Dependence}

Let us discuss the numerical results for the static spin
susceptibility at ${\bf q}={\bf Q}$, $\chi_s$. 
The dependence on $\delta$ of $\chi_s$ is depicted
in Fig.~\ref{fig:chisdoping1}. 
Because of the nesting properties of the Fermi surface, the AF
correlation develops near the half-filling and results in the
enhancement of $\chi_s$. 
We also show the results for the case in the presence of the next 
nearest-neighbor hopping $t'$ in Figs.~\ref{fig:chisdoping2} and 
\ref{fig:chisdoping3}. 
In both cases of $t'=0.2$ and $t'=-0.2$, the nesting properties of the 
Fermi surface is much weakened. 
Thus the enhancement of AF correlation is reduced compared with the 
case for $t'=0$, but there is no qualitative difference in the doping
dependence between these cases. 

\subsection{Temperature Dependence: Spin-gap-like Behavior} 

In this subsection, we investigate the temperature dependence of 
$\chi_s$. 
The results for the case with only the nearest-neighbor hopping are 
shown in Fig.~\ref{fig:chistemp1}. 
The electron number is chosen as $n=0.94$. 
In the high-temperature region, $\chi_s$ follows the Curie-Weiss-like 
behavior. 
On the other hand, in the low-temperature region, the spin-gap-like 
behavior manifests. 
Since it is observed even in the case of $U=0$, this spin-gap-like
behavior is essentially due to the band effect: 
In the non-half-filled case, VHS's at ${\bf k}=(\pm\pi, \pm\pi)$ are
not on the Fermi surface and then the contributions from VHS to the
spin susceptibility decrease as the temperature is lowered. 
Consequently, $\chi_s$ shows the spin-gap-like behavior in the
low-temperature region. 
Although the spin-gap-like behavior is very small for $U=0$ near the 
half-filling, it is much enhanced due to the correlation effect 
and results in the large spin-gap-like behavior as shown in 
Fig.~\ref{fig:chistemp1}. 
Note that such enhancement is not brought about by the speciality of 
RPA, because we also obtain the robust spin-gap-like structure even 
with the vertex corrections beyond RPA. 

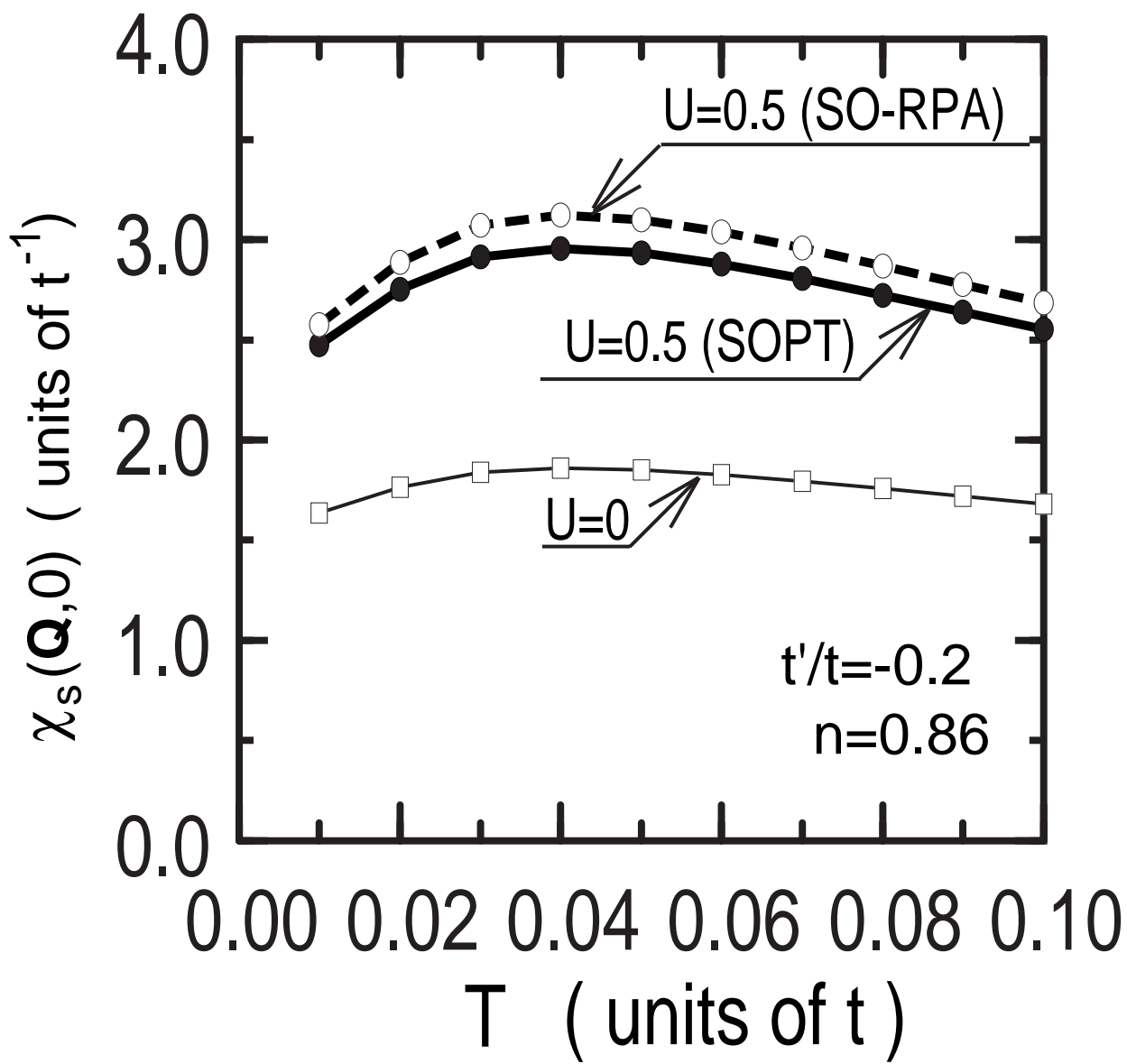
\begin{figure}
  \caption{Spin susceptibility plotted as a function of the 
    temperature $T$ for $t'=-0.2t$ and $n=0.86$. 
    The results are shown in the same lines and symbols as in Fig.~8.}
  \label{fig:chistemp4}
\end{figure}

When we introduce the next nearest-neighbor hopping, the spin-gap-like
behavior disappears as shown in Figs.~\ref{fig:chistemp2} and 
\ref{fig:chistemp3} for $t'=0.2$ and $t'=-0.2$, respectively, with 
$n=0.94$. 
The next nearest-neighbor hopping changes band structure and the
effect of VHS becomes too small to give rise the spin-gap behavior in 
these cases. 
For the case of $t'=-0.2$, the large enhancement in the
low-temperature region can be understood as the effect of VHS, 
because the location of VHS is very close to the Fermi surface. 
At the lower temperature than $0.01t$, it is expected that the
spin-gap behavior may manifest, but we do not carry out the 
calculation in this temperature region because of the lack of the 
reliability of our calculation on a $64 \times 64$ lattice as well as
the limitation of computer memory. 
Instead of calculating $\chi_s$ at lower temperature, we obtain the
temperature dependence of $\chi_s$ at $n=0.86$. 
The results are shown in Fig.~\ref{fig:chistemp4}. 
Since the difference between the position of VHS and that of the Fermi 
level is $0.05t$ at this filling, the spin-gap-like behavior is
observed even in the present temperature range in contrast to the case 
of $n=0.94$. 
Thus our spin-gap-behavior is originating from VHS, not the nesting
property of the Fermi surface. 

\section{Summary and Discussion}

We have investigated the doping and temperature dependence of the 
charge and spin susceptibilities by using SOPT. 
The charge susceptibility is much suppressed due to the correlation
effect through the vertex corrections in the vicinity of the
half-filling for $U$ comparable to $t$. 
It has been explicitly shown that the vertex corrections beyond RPA 
are indispensable in order to obtain this behavior. 
The suppression of the charge susceptibility near the half-filling 
indicates a precursor of the formation of the Mott-Hubbard gap. 
Such a behavior of the charge susceptibility will be qualitatively
correct as long as the system is kept in the Fermi-liquid state, 
but we cannot perfectly deny the possibility that the charge 
susceptibility jumps abruptly from a finite value to zero or diverges
at the half-filling if the system turns to be in a non-Fermi-liquid
state due to the speciality of two-dimensional systems in combination
with the strong correlation effect. 
In this sense, we need further studies beyond perturbative calculation 
in order to understand the behavior of the charge susceptibility near  
MIT in the two-dimensional Hubbard model. 

In the study of the spin susceptibility, we have obtained the
spin-gap-like behavior not only the case for $t'=0$ 
but also the case with non-zero next nearest-neighbor hopping. 
The origin of our spin-gap-like behavior is ascribed to the 
band effect such as VHS in combination with the enhancement effect 
due to electron correlation: 
In the static spin susceptibility at ${\bf q}=(\pi,\pi)$, VHS gives
rise to a small spin-gap-like structure, which will be much enhanced
by correlation effects. 
We have emphasized that such an enhancement is not brought about by
the speciality of RPA. 
Even if we take into account the diagrams for the vertex corrections
beyond RPA, the same structure has been obtained. 
Those vertex corrections are important in the two-dimensional model
in which quantum and thermal fluctuations destroy any long range
order. 

Finally, we give a brief comment on the spin gap in NMR experiments 
on HTSC.
\cite{kitaoka1,kitaoka2,yasuoka1,warren,takigawa,zimm,yasuoka2,zheng,wals}
It has been argued that the origin of the spin-gap-like behavior 
observed in those experiments is the condensation of paired spinons in
the RVB picture \cite{fuku} or the bi-layer coupling in the
Fermi-liquid picture. \cite{millis} 
As we have seen in this paper, even in the Fermi-liquid theory without 
bi-layer coupling, the spin-gap-like behavior has been 
reproduced by the effect of VHS combined with the enhancement effect 
due to electron correlation. 
Although it is difficult to adjust quantitively the size of the gap
and the band structure in our calculation to the experimental results,
there are some indirect evidences which imply the relevance of VHS 
to the spin-gap behavior. 
In the NMR experimental result on ${\rm YB_2Cu_3O_{7-\delta}}$
\cite{takigawa}, it can be seen that the spin-gap behavior becomes
weak as $\delta$ approaches zero. 
It has been pointed out that the DOS of ${\rm YB_2Cu_3O_{7-\delta}}$
has a peak near the Fermi level due to VHS for $\delta=0$ and this 
peak runs away from the Fermi level as $\delta$ increases\cite{tsu}. 
Thus it may be plausible to attribute the spin-gap-like behavior of
this material to VHS in DOS enhanced due to strong electron 
correlation. 

\acknowledgements

The authors are grateful to K. Yamada, Y. Takada, N. Kawakami,
M. Imada, H. Fukuyama, and N. Furukawa for useful conversations. 
This work has been supported by a Grant-in-Aid for Scientific Research 
on Priority Area ``Anomalous Metallic State near the Mott Transition'' 
(07237102) from the Ministry of Education, Science, Sports, and
Culture, Japan. 


\end{multicols}


\narrowtext

\begin{figure}
  \caption{Diagrams for $\chi^{\sigma \sigma'}$ within second order
    with respect to the on-site Coulomb interaction $U$. 
    The solid and the dashed lines, respectively, indicate the bare
    Green's function $G_0$ and $U$. 
    Note that the Hartree term is included in the $G_0$-line.} 
  \label{fig:diagrams}
\end{figure}

\begin{figure}
  \caption{Charge susceptibility plotted as a function of the
    hole-doping $\delta$ for $t'=0$ and $T=0.04t$. 
    Open squares linked by thin solid line indicates the results for 
    $U=0$. 
    The results for SO-RPA and SOPT for $U=t$ are denoted 
    by open circles connected by broken line and solid circles linked
    by thick line, respectively.}
  \label{fig:chicdoping1}
\end{figure}

\begin{figure}
  \caption{Charge susceptibility plotted as a function of the
    hole-doping $\delta$ for $t'=0.2t$ and $T=0.04t$. 
    The results are shown in the same lines and symbols as in Fig.~2.}
  \label{fig:chicdoping2}
\end{figure}

\begin{figure}
  \caption{Charge susceptibility plotted as a function of the
    hole-doping $\delta$ for $t'=-0.2t$ and $T=0.04t$. 
    The results are shown in the same lines and symbols as in Fig.~2.}
  \label{fig:chicdoping3}
\end{figure}

\begin{figure}
  \caption{Spin susceptibility as a function of the 
    hole-doping $\delta$ for $t'=0$ and $T=0.04t$. 
    Open squares linked by this solid line indicates the results for 
    $U=0$. 
    The results for SO-RPA and SOPT for $U=0.5t$ are denoted by open
    circles connected by broken line and solid circles linked by thick
    line, respectively.} 
  \label{fig:chisdoping1}
\end{figure}

\begin{figure}
  \caption{Spin susceptibility as a function of the hole-doping
    $\delta$ for $t'=0.2t$ and $T=0.04t$. 
    The results are shown in the same lines and symbols as in Fig.~5.}
  \label{fig:chisdoping2}
\end{figure}

\begin{figure}
  \caption{Spin susceptibility as a function of the hole-doping
    $\delta$ for $t'=-0.2t$ and $T=0.04t$. 
    The results are shown in the same lines and symbols as in Fig.~5.}
  \label{fig:chisdoping3}
\end{figure}

\begin{figure}
  \caption{Spin susceptibility plotted as a function of the 
    temperature $T$ for $t'=0$ and $n=0.94$. 
    Open squares linked by this solid line indicates the results for
    $U=0$. 
    The results for SO-RPA and SOPT for $U=0.5t$ are denoted 
    by open circles connected by broken line and solid circles linked
    by thick line, respectively. }
  \label{fig:chistemp1}
\end{figure}

\begin{figure}
  \caption{Spin susceptibility plotted as a function of the 
    temperature $T$ for $t'=0.2t$ and $n=0.94$.
    The results are shown in the same lines and symbols as in Fig.~8.}
  \label{fig:chistemp2}
\end{figure}

\begin{figure}
  \caption{Spin susceptibility plotted as a function of the 
    temperature $T$ for $t'=-0.2t$ and $n=0.94$. 
    The results are shown in the same lines and symbols as in Fig.~8.}
  \label{fig:chistemp3}
\end{figure}

\end{document}